\documentclass[aps,prb,twocolumn,showpacs,showkeys,floatfix]{revtex4}

\usepackage{graphicx,color}

\newcommand{\jwj}[1]{\textcolor{red}{#1}}

\graphicspath{{figs/}}
\begin{document}
\title{Phonon Bandgap Engineering of Strained Monolayer MoS$_{2}$}
\author{Jin-Wu Jiang}
    \altaffiliation{Corresponding author: jwjiang5918@hotmail.com}
    \affiliation{Shanghai Institute of Applied Mathematics and Mechanics, Shanghai Key Laboratory of Mechanics in Energy Engineering, Shanghai University, Shanghai 200072, People's Republic of China}

\date{\today}
\begin{abstract}
The phonon band structure of monolayer MoS$_{2}$ is characteristic for a large energy gap between acoustic and optical branches, which protects the vibration of acoustic modes from being scattered by optical phonon modes. Therefore, the phonon bandgap engineering is of practical significance for the manipulation of phonon-related mechanical or thermal properties in monolayer MoS$_{2}$. We perform both phonon analysis and molecular dynamics simulations to investigate the tension effect on the phonon bandgap and the compression induced instability of the monolayer MoS$_{2}$. Our key finding is that the phonon bandgap can be narrowed by the uniaxial tension, and is completely closed at $\epsilon=0.145$; while the biaxial tension only has limited effect on the phonon bandgap. We also demonstrate the compression induced buckling for the monolayer MoS$_{2}$. The critical strain for buckling is extracted from the band structure analysis of the flexure mode in the monolayer MoS$_{2}$ and is further verified by molecular dynamics simulations and the Euler buckling theory. Our study illustrates the uniaxial tension as an efficient method for manipulating the phonon bandgap of the monolayer MoS$_{2}$, while the biaxial compression as a powerful tool to intrigue buckling in the monolayer MoS$_{2}$.

\end{abstract}

\maketitle
\pagebreak

\section{Introduction}

Molybdenum Disulphide (MoS$_{2}$) is a semiconductor with a bulk bandgap above 1.2~{eV},\cite{KamKK} which can be further manipulated by changing its thickness,\cite{MakKF} or through application of mechanical strain.\cite{FengJ2012npho,LuP2012pccp,ConleyHJ,CheiwchanchamnangijT2013prb} This finite bandgap is a key reason for the excitement surrounding MoS$_{2}$ as compared to graphene as graphene has a zero bandgap.\cite{NovoselovKS2005nat}  Because of its direct bandgap and also its well-known properties as a lubricant, MoS$_{2}$ has attracted considerable attention in recent years.\cite{WangQH2012nn,ChhowallaM,Butler2013acsnn,Xu2013cr,JariwalaD2014acsnn} For example, Radisavljevic et al.\cite{RadisavljevicB2011nn} demonstrated the application of monolayer MoS$_{2}$ as a good transistor. The strain and the electronic noise effects were found to be important for the monolayer MoS$_{2}$ transistor.\cite{ConleyHJ,SangwanVK,Ghorbani-AslM,CheiwchanchamnangijT}  Several recent works have addressed the thermal transport properties of monolayer MoS$_{2}$ in both ballistic and diffusive transport regimes.\cite{HuangW,JiangJW2013mos2,VarshneyV,JiangJW2013sw} The mechanical behavior of monolayer MoS$_{2}$ has been investigated.\cite{BertolazziS,CooperRC2013prb1,CooperRC2013prb2,JiangJW2013sw} Quite recently, we derived an analytic formula for the elastic bending modulus of the monolayer MoS$_{2}$, where the importance of the finite thickness effect was revealed.\cite{JiangJW2013bend} We have also shown that the MoS$_{2}$ resonator has much higher quality factor than the graphene resonator.\cite{JiangJW2013mos2resonator}

The electronic properties in the monolayer MoS$_{2}$ are dominated by its electronic band structure (particularly its large bandgap). The thermal and mechanical properties in the monolayer MoS$_{2}$ are mainly determined by its phonon band structure (i.e phonon spectrum or phonon dispersion). A distinct feature in the phonon band structure of the monolayer MoS$_{2}$ is the big energy gap between acoustic branches and optical branches, which has been confirmed by the experimental measurement\cite{WakabayashiN} and various theoretical calculations.\cite{JimenezSS,DamnjanovicM2008mmp,SanchezAM} The bandgap separates the acoustic phonon modes from the optical modes in the monolayer MoS$_{2}$. As a result, the vibration of the acoustic modes is difficult to be scattered by the optical modes, because it is difficult to satisfy the energy conservation law owning to the big bandgap. In other words, the acoustic vibration is preserved by the phonon bandgap from being interrupted by the optical modes. Due to this preserving mechanism, it has been shown that the MoS$_{2}$ resonator has higher quality factor than the graphene resonator.\cite{JiangJW2013mos2resonator} The phonon bandgap induced preserving mechanism is expected to play important roles in other thermal and mechanical properties in the monolayer MoS$_{2}$, \jwj{such as the thermal conductivity, thermal expansion, and the nano-mechanical resonator}. Because of this, the engineering of the phonon bandgap becomes practically important, since it can be used to efficiently manipulate various thermal and mechanical properties in the monolayer MoS$_{2}$.

In graphene, the strain engineering has been proved to be an effective approach to modify its various properties.\cite{LiuF2007prb,GuiG2008prb,SunL2008jcp,NiZH2008acsnano,MohiuddinTMG2009prb,PereiraVM2009prl,GuineaF2010natp,PellegrinoFMD2010prb,LevyN2010sci,JiangJW2011negf,WeiN,KumarSB2012nl} In particular, lots of works have shown the electronic bandgap opening in the strained graphene. For monolayer MoS$_{2}$, the strain is also able to manipulate its electronic bandgap efficiently.\cite{FengJ2012npho,LuP2012pccp,EmilioS2012nnr,ConleyHJ,CheiwchanchamnangijT2013prb} So far, the study of the strain effect on the phonon band structure in monolayer MoS$_{2}$ is still lacking, and is thus the focus of the present work.

In this paper, we study the phonon band structure in the monolayer MoS$_{2}$ with uniaxial or biaxial strains, using both phonon band structure calculation and molecular dynamics (MD) simulations. The uniaxial tension is found to be more effective than the biaxial tension for the bandgap manipulation in the monolayer MoS$_{2}$. The phonon bandgap between acoustic and optical phonon branches can be closed by the uniaxial tension $\epsilon=0.145$, while the biaxial tension has only limited effect on the phonon bandgap. It implies that the uniaxial tension can have significant effect on the lattice dynamics properties of the monolayer MoS$_{2}$, where the phonon bandgap plays an important role. In the compressed monolayer MoS$_{2}$, the long wave flexure phonon modes have negative frequency, which indicates the instability of the quasi-two-dimensional monolayer MoS$_{2}$. The band structure analysis for the flexure mode gives the relationship between the critical length and the critical compression, which is further confirmed by MD simulations and the Euler buckling theory.

The rest of the present paper is organized as follows. In Sec.~II, we present the tension effect on the phonon bandgap in the monolayer MoS$_{2}$. Sec.~III is devoted to the compression induced buckling of the monolayer MoS$_{2}$. The paper ends with a brief summary in Sec.IV.

\begin{figure}[htpb]
  \begin{center}
    \scalebox{0.9}[0.9]{\includegraphics[width=8cm]{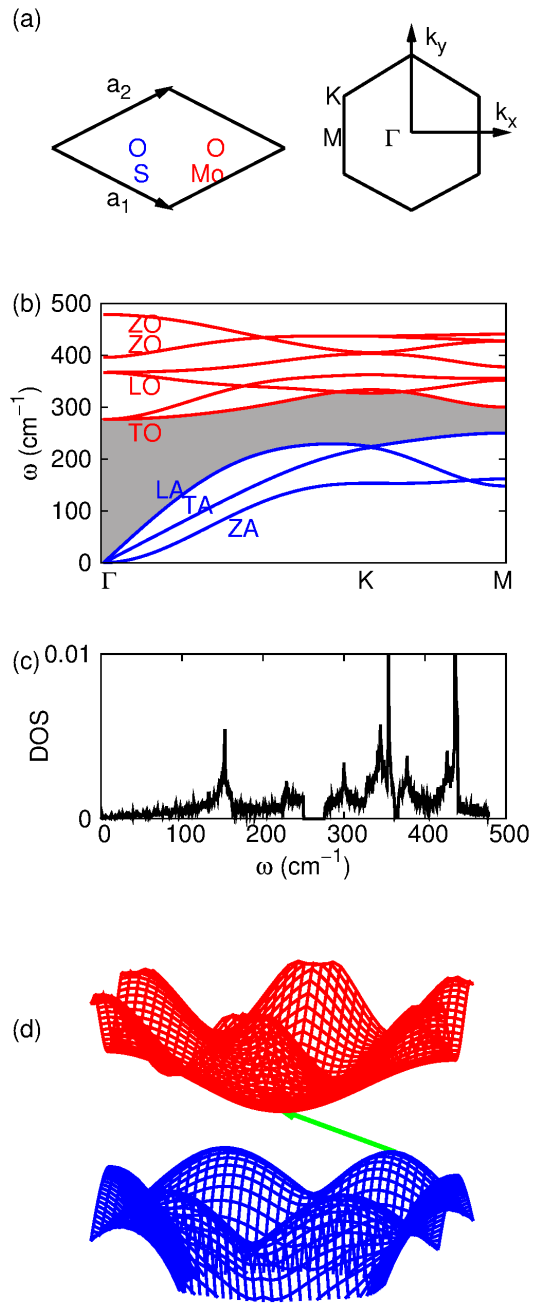}}
  \end{center}
  \caption{(Color online) Phonon band structure of monolayer MoS$_{2}$. (a) Left: primitive unit cell for the MoS$_{2}$ hexagonal lattice. Right: The first Brillouin zone for the hexagonal lattice. $\Gamma$, K, and M are three high symmetry points. (b) Phonon band structure along high symmetry lines in the first Brillouin zone. Note the bandgap (gray area) between acoustic and optical branches. (c) Phonon DOS in the monolayer MoS$_{2}$. Note the zero DOS around 260~{cm$^{-1}$}, implying a bandgap in this frequency range. (d) Three dimensional phonon band structure for the two branches around the bandgap, i.e the highest-frequency acoustic branch and the lowest-frequency optical branch. The green arrow indicates the bandgap.}
  \label{fig_dispersion}
\end{figure}

\begin{figure}[htpb]
  \begin{center}
    \scalebox{1}[1]{\includegraphics[width=8cm]{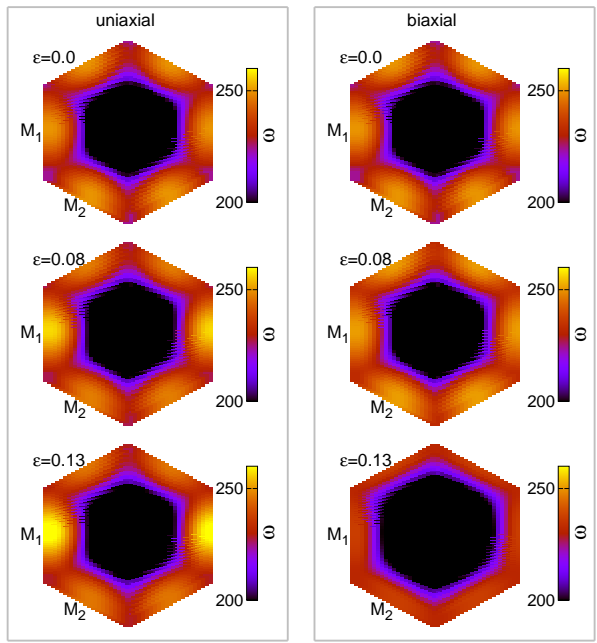}}
  \end{center}
  \caption{(Color online) The highest-frequency acoustic phonon spectrum for monolayer MoS$_{2}$ under tension. The highest frequency locates at M point in the Boillouin zone. Left: uniaxial tension effect. The degeneracy between frequencies for M$_{1}$ and M$_{2}$ is removed, due to the anisotropic tension style. Right: biaxial tension effect. The degeneracy between frequencies for M$_{1}$ and M$_{2}$ is maintained owning to the isotropic tension style. Note the abrupt decrease of the frequency at M point at biaxial tension $\epsilon=0.13$.}
  \label{fig_acoustic_strain}
\end{figure}

\begin{figure}[htpb]
  \begin{center}
    \scalebox{1}[1]{\includegraphics[width=8cm]{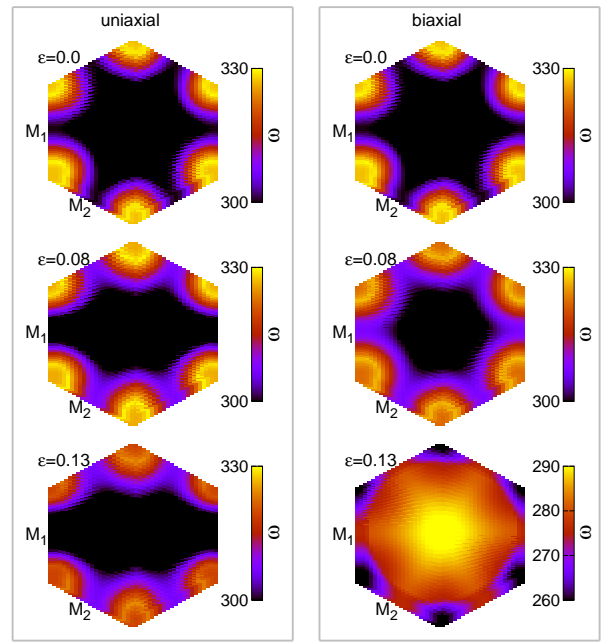}}
  \end{center}
  \caption{(Color online) The lowest-frequency optical phonon spectrum for monolayer MoS$_{2}$ under tension. The lowest frequency locates at the Brillouin zone center for monolayer MoS$_{2}$ without tension. Left: uniaxial tension effect. The degeneracy between frequencies for M$_{1}$ and M$_{2}$ is removed, due to the anisotropic tension style. Right: biaxial tension effect. The degeneracy between frequencies for M$_{1}$ and M$_{2}$ is maintained owning to the isotropic tension style. Note the lowest frequency position changes from $\Gamma$ to K point in the Brillouin zone at biaxial tension $\epsilon=0.13$.}
  \label{fig_optical_strain}
\end{figure}

\begin{figure}[htpb]
  \begin{center}
    \scalebox{1}[1]{\includegraphics[width=8cm]{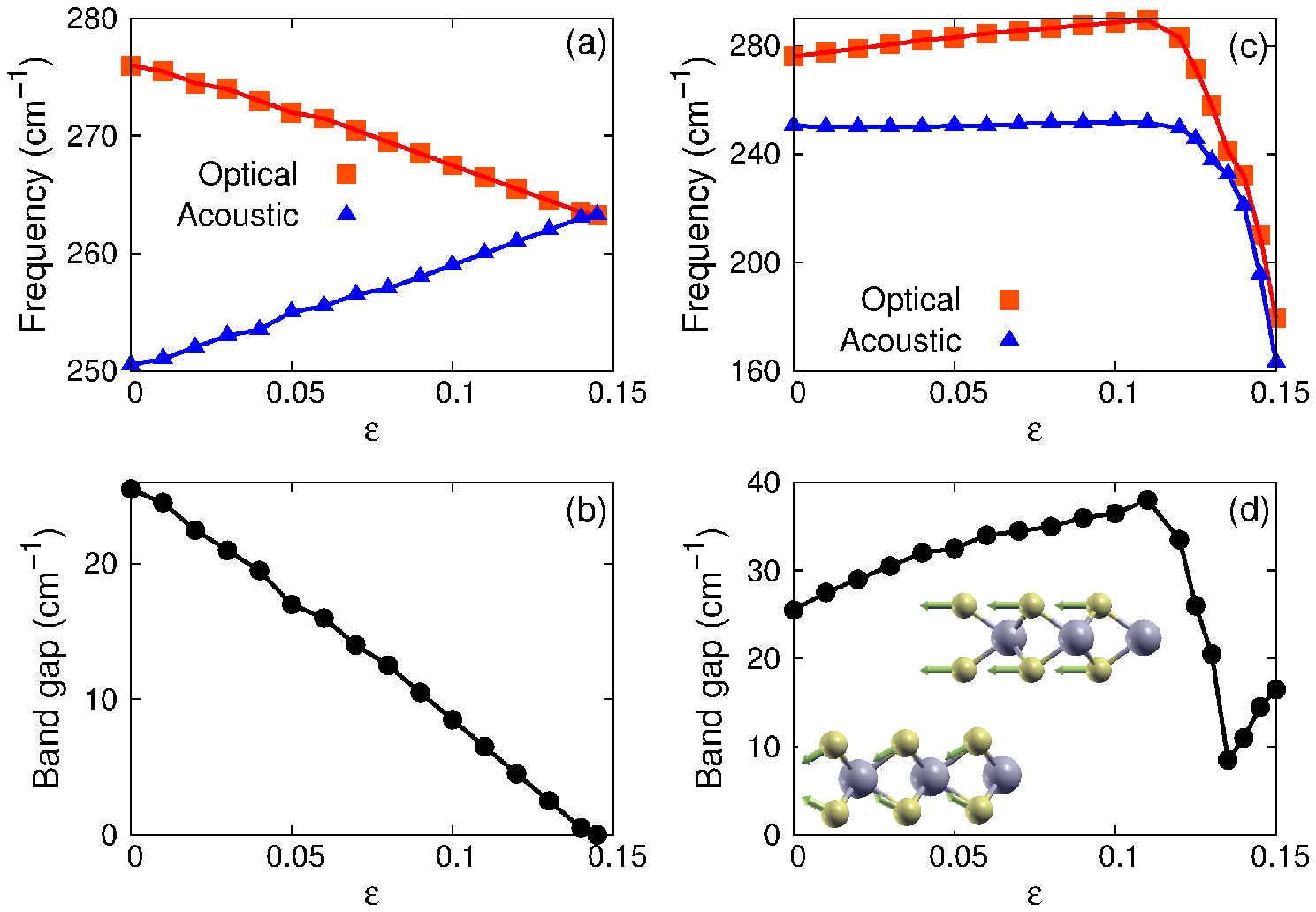}}
  \end{center}
  \caption{(Color online) Tension effect on the phonon bandgap. (a) Uniaxial tension effect on the lowest frequency in the lowest-frequency optical branch and the highest frequency in the highest-frequency acoustic branch. (b) Bandgap versus the uniaixal tension. The bandgap is closed at uniaxial tension $\epsilon=0.145$. (c) Biaxial tension effect on the lowest frequency in the lowest-frequency optical branch and the highest frequency in the highest-frequency acoustic branch. (d) Bandgap versus the biaixal tension, where insets show the vibration morphology of the highest-frequency acoustic mode and the lowest-frequency optical mode.}
  \label{fig_bandgap}
\end{figure}

\section{Tension effect on the phonon band structure}
Fig.~\ref{fig_dispersion} shows the phonon band structure in the monolayer MoS$_{2}$ without strain engineering. The monolayer MoS$_{2}$ has a hexagonal honeycomb lattice structure. Fig.~\ref{fig_dispersion}~(a) left shows the primitive unit vector of the honeycomb lattice. There are one Mo atom and two S atoms within the primitive unit cell. These two S atoms fall into the same lattice site in the honeycomb structure, and the Mo atom takes up the other lattice site. $\vec{a}_{1}$ and $\vec{a}_{2}$ are the primitive lattice vectors in the honeycomb lattice. The lattice constant is $|\vec{a}_{1}|=|\vec{a}_{2}|=2.14$~{\AA}. Fig.~\ref{fig_dispersion}~(a) right shows the first Brillouin zone of the honeycomb lattice structure. Three high symmetry points are denoted by $\Gamma$, K, and M. The phonon band structure is calculated from the lattice dynamics properties package GULP.\cite{gulp} The interaction within MoS$_{2}$ is described by the Stillinger-Weber potential.\cite{JiangJW2013sw}

Fig.~\ref{fig_dispersion}~(b) shows the phonon band structure of the monolayer MoS$_{2}$ along the high symmetry line $\Gamma$KM in the first Brillouin zone. There are totally nine branches in the phonon band structure corresponding to the nine degrees of freedom in the primitive unit cell of the monolayer MoS$_{2}$. More specifically, there are three acoustic branches, i.e the flexure mode (ZA) branch, the transverse acoustic (TA) branch, and the longitudinal acoustic (LA) branch. There are six optical branches, i.e two z-direction optical (ZO) branches, two transverse optical (TO) branches, and two longitudinal optical (LO) branches. A distinct feature in this phonon band structure is the large energy gap (gray area) between the three acoustic branches and the six optical branches. The acoustic and optical branches are separated by this bandgap. As a result, it is rather difficult to satisfy the energy conservation law during the phonon-phonon scattering between acoustic and optical modes. In other words, the acoustic phonon modes are difficult to interact with the optical phonon modes. Hence, the vibration of the acoustic phonon modes can be preserved by the bandgap from being interrupted by the optical phonon modes. Owning to the bandgap induced preserving mechanism, it has been shown that the monolayer MoS$_{2}$ resonator has higher quality factor than the graphene resonator.\cite{JiangJW2013mos2resonator}

Fig.~\ref{fig_dispersion}~(c) shows the phonon density of state (DOS) in the monolayer MoS$_{2}$. The DOS is calculated from the whole phonon band structure in the first Brillouin zone with $201\times201\times1$ k-sampling. There is a zero DOS region around 260~{cm$^{-1}$}, which corresponds to the phonon bandgap shown in panel (b). Fig.~\ref{fig_dispersion}~(d) shows the three-dimensional plot for the two phonon branches at the bandgap, i.e the highest-frequency acoustic branch and the lowest-frequency optical branch. In the following, we will refer to the highest-frequency acoustic branch as lower band, and the lowest-frequency optical branch as upper band. For the monolayer MoS$_{2}$ without strain, the minimum frequency in the upper band locates at the $\Gamma$ point in the first Brillouin zone, and the maximum frequency in the lower band locates at the six M points in the first Brillouin zone. Hence, the monolayer MoS$_{2}$ can be regarded as an `indirect semiconductor'.

We study the tension effect on the phonon band structure of the monolayer MoS$_{2}$. We focus on the two branches at the bandgap shown in Fig.~\ref{fig_dispersion}~(d). The uniaxial and biaxial tensions are comparatively investigated. The primitive unit cell is optimized to the energy minimum configuration prior tension. The uniaxial tension is applied to the optimized unit cell in the $x$ direction. After the uniaxial tension is applied, the unit cell is frozen in its $x$ direction and is allowed to relax in the $y$ direction. For the biaxial tension, the pre-optimized unit cell is stretched in both $x$ and $y$ directions and no further relaxation is allowed for the unit cell. In both uniaxial and biaxial tensions, all three atoms within the unit cell are allowed to be fully relaxed.

Fig.~\ref{fig_acoustic_strain} shows the tension effect on the lower band. In the ideal monolayer MoS$_{2}$ without tension, there is a six-fold symmetry in the phonon band structure. The maximum frequency locates at the six M points in the Brillouin zone. Under uniaxial tension, the six-fold symmetry is removed due to the asymmetric strain style. The maximum frequency locates at the two M$_{1}$ points in the uniaxially tensile monolayer MoS$_{2}$. For biaxial tension, the six-fold symmetry in the phonon band structure is maintained owning to the symmetric strain type. The highest-frequency still locates at the M points in the biaxially tensile monolayer MoS$_{2}$. However, for biaxial tension $\epsilon=0.13$, there is an abrupt decrease in the value of the maximum frequency in the phonon band structure. This implies some structure transition in the system.

Fig.~\ref{fig_optical_strain} shows the tension effect on the upper band. The six-fold symmetry in the band structure is broken by the asymmetric uniaixal tension. For uniaxial tension, the location of the minimum frequency remains in the $\Gamma$ point of the first Brillouin zone. Under biaxial tension, the band structure keeps the six-fold symmetry. It is quite interesting that the minimum frequency location shifts from the $\Gamma$ to the K point at large biaxial tension $\epsilon=0.13$, which implies a qualitative change in the phonon band structure induced by the structure transition.

\begin{figure}[htpb]
  \begin{center}
    \scalebox{1}[1]{\includegraphics[width=8cm]{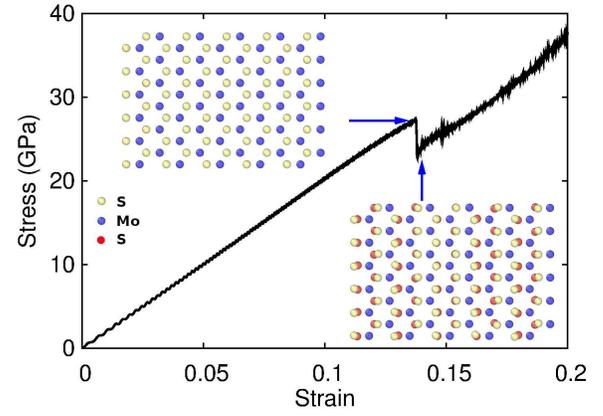}}
  \end{center}
  \caption{(Color online) The stress-strain relation during the biaxial tension of the monolayer MoS$_{2}$ of dimension {30\AA$\times$20\AA} at 1.0~K. A sudden jump in the stress at $\epsilon=0.137$ discloses a structure transition. Insets show the top view of the monolayer MoS$_{2}$ just before or after the structure transition.}
  \label{fig_stress_strain_tension}
\end{figure}

\begin{figure*}[htpb]
  \begin{center}
    \scalebox{0.9}[0.9]{\includegraphics[width=\textwidth]{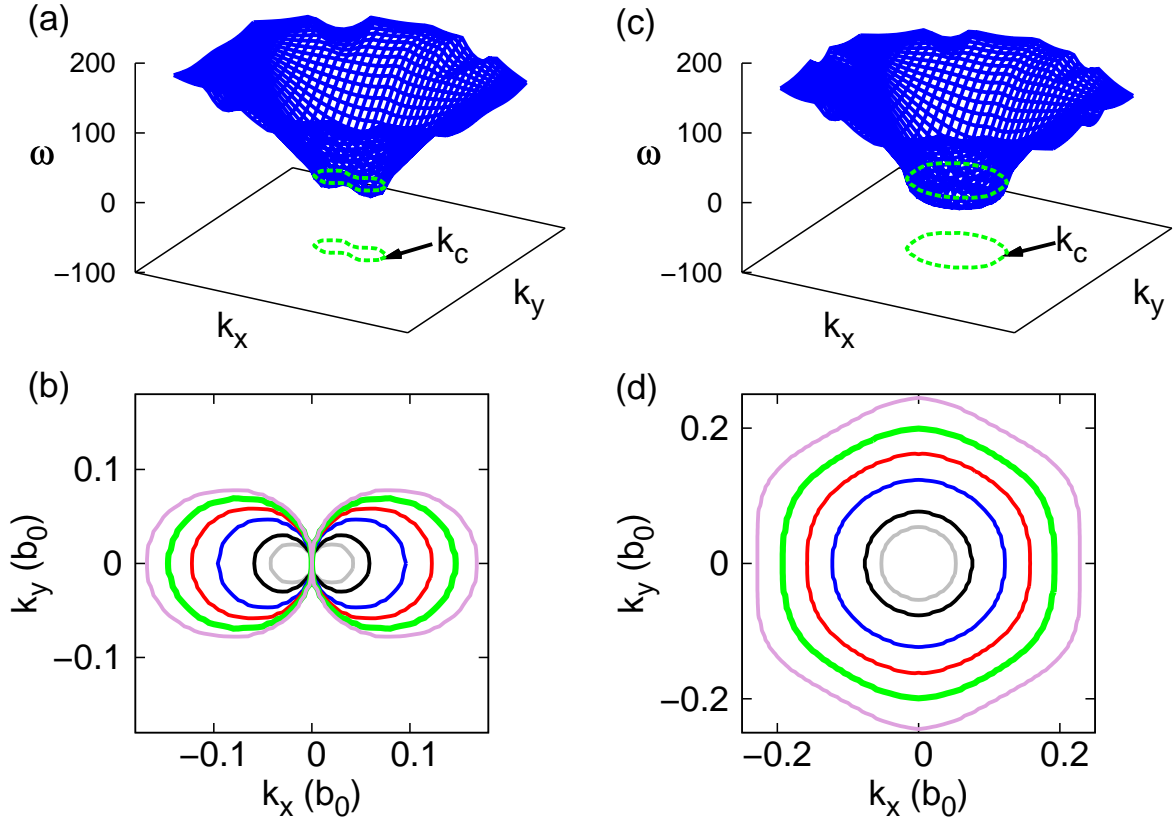}}
  \end{center}
  \caption{(Color online) Compression effect on the flexure mode branch of monolayer MoS$_{2}$. (a) The phonon band structure of the flexure mode under uniaxial compression $\epsilon=-0.08$. The glass-shaped circle (green online) is the contour for $\omega=0$. The frequency is negative for wave vectors enclosed by the contour. $k_{c}$ is the maximum wave vector on the contour. (b) From inside to outside, the expansion of the contour for $\omega=0$ with increasing uniaxial compression $\epsilon=-0.01$, -0.02, -0.05, -0.08, -0.11, and -0.14. (c) The phonon band structure of the flexure mode under biaxial compression $\epsilon=-0.08$. The contour for $\omega=0$ has six-folded symmetry owning to the isotropic compression style. (d) From inside to outside, the expansion of the contour for $\omega=0$ with increasing biaxial compression $\epsilon=-0.01$, -0.02, -0.05, -0.08, -0.11, and -0.14.}
  \label{fig_dispersion_strain}
\end{figure*}

The tension effect on the phonon bandgap is shown in Fig.~\ref{fig_bandgap}. Panel (a) shows the uniaixal tension effect on the minimum frequency (red squares) from the upper band, and the maximum frequency (blue triangles) from lower band. With the increase of the uniaixal tension, the minimum frequency decreases monotonically and the maximum frequency increases monotonically. As a result, the phonon bandgap can be narrowed by applying stronger uniaxial tension as shown in Fig.~\ref{fig_bandgap}~(b). The phonon bandgap is completely closed for uniaxial tension stronger than 0.145. This results predicts that the thermal or mechanical properties of the monolayer MoS$_{2}$ should behavior quite differently when it is stretched uniaxially for a strain above 0.145. \jwj{It should be noted that uniaxial tension of $\epsilon = 0.145$ is in a reasonable strain range for the monolayer MoS$_{2}$. It is because the nano-indentation measurement found that breaking occurs at an effective strain between 6\% and 11\% for the ultra thin MoS$_{2}$.\cite{BertolazziS} Considering tip-induced stress concentration and the unavoidable defect in the experiment sample, the theoretical ultimate strain for monolayer MoS$_{2}$ should be above higher. Indeed, first-principles calculations have predicted an ultimate strain above 20\% for the monolayer MoS$_{2}$ during tension.\cite{CooperRC2013prb1,TaoP2014jap}}

 Panels (c) and (d) show that the biaxial tension is not an effective approach to manipulate the phonon bandgap in monolayer MoS$_{2}$. Both the minimum frequency and the maximum frequency are almost not affected by the biaxial tension for $\epsilon<0.13$. There is an abrupt decrease in the minimum and maximum frequencies, when the system is biaxially stretched for $\epsilon>0.13$.

We discuss more details for the biaxial tension induced structure transition. The insets in Fig.~\ref{fig_bandgap}~(d) show the vibration morphology of the highest-frequency acoustic mode at the K point in the Brillouin zone and the lowest-frequency optical mode at the M point in the Brillouin zone. These two modes are calculated for the monolayer MoS$_{2}$ stretched biaxially with $\epsilon=0.13$. In both modes, the two S atomic layers vibrate mainly in the xy plane, while the middle Mo atomic layer does not vibrate. A sudden decrease in the frequency of these two modes indicates a structure transition under the biaxial tension. It also implies that, during the transition, the structure evolves in a way similar as the vibration morphology of these two modes; i.e the outer two S atomic layers should shift with respect to the inner Mo atomic layer.

To further explore the structure transition, we perform the MD simulation for the tensile behavior of a monolayer MoS$_{2}$ with dimension {30\AA$\times$20\AA}. All MDs simulations in this work are performed using the publicly available simulation code LAMMPS~\cite{PlimptonSJ,Lammps}, while the OVITO package was used for visualization in this section~\cite{ovito}. The standard Newton equations of motion are integrated in time using the velocity Verlet algorithm with a time step of 1 fs. The interaction within MoS$_{2}$ is described by the Stillinger-Weber potential.\cite{JiangJW2013sw} MD simulations are run at a very low temperature (T=1.0~K), which is to keep consistent with the above phonon band structure calculation, which does not consider the temperature effect either. Periodic boundary conditions are applied in the two in-plane directions, and the free boundary condition is applied in the out-of-plane direction.

Fig.~\ref{fig_stress_strain_tension} shows the stress-strain relation for the biaxial tension of the monolayer MoS$_{2}$. Two insets show the configuration just before and after the structure transition. The structure transition happens at $\epsilon=0.137$, which is in good agreement with the above phonon analysis. In the new structure after the transition, the two outer S atomic layers are indeed shifted with respect to the inner Mo atomic layer, which is again consistent with the phonon analysis. \jwj{The shift of outer S atomic layers is also observed in the experiment, where the MoS$_{2}$ becomes metallic after the structure transition.\cite{GokiE2012acsn,LinYC2013}}

\begin{figure}[htpb]
  \begin{center}
    \scalebox{1}[1]{\includegraphics[width=8cm]{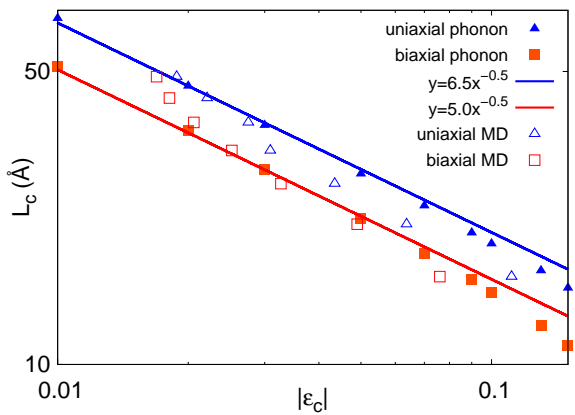}}
  \end{center}
  \caption{(Color online) The relationship between the critical length ($L_{c}$) and the critical compression ($\epsilon_{c}$) from lattice dynamics calculations and MD simulations. Fill points are prediction from the lattice dynamics calculation of the flexure mode for the monolayer MoS$_{2}$  compressed by $\epsilon_{c}$, where the critical length $L_{c}=2\pi/k_{c}$ and $k_{c}$ is the wave vector depicted in Fig.~(a) or (c). Lines are fitting to data from the lattice dynamics calculation. Open points are results from MD simulations, where the monolayer MoS$_{2}$ with length $L_{c}$ becomes unstable when it is compressed by a compression stronger than the critical value (i.e $|\epsilon|>|\epsilon_{c}|$).}
  \label{fig_lc}
\end{figure}

\section{Compression induced instability}
In the above we have studied the tension effect on the phonon band structure of the monolayer MoS$_{2}$. The rest of this paper will be devoted to the investigation of the compression effect on the monolayer MoS$_{2}$. Under compression, the quasi-two-dimensional monolayer MoS$_{2}$ will be unstable (via buckling) if the applied compression is strong enough. Owning to the flexibility of the two-dimensional structure, a strong compression is most likely to induce the out-of-plane deformation in the monolayer MoS$_{2}$, i.e the formation of ripples, i.e buckling.

This compression induced instability can be well captured by the phonon band structure. In terms of phonon modes, the ripple formation is actually equivalent to the emergence of the flexure modes in the monolayer MoS$_{2}$. It is because the sinuous shape of the ripple is the same as the vibration morphology of the flexure mode. We thus can investigate the compression induced instability of the monolayer MoS$_{2}$ by examining the compression effect on its phonon band structure (particularly the flexure mode).

Fig.~\ref{fig_dispersion_strain} shows the band structure for the flexure mode in monolayer MoS$_{2}$ under uniaxial or biaxial compressions. Panel (a) shows that the frequency of the long wave (i.e around $\Gamma$ point) flexure phonons become negative when the monolayer MoS$_{2}$ is uniaxially compressed by $\epsilon=-0.08$. This negative frequency in the flexure mode indicates that the two-dimensional structure is not stable if it is deformed in the way similar as the vibration morphology of the flexure mode. The eigen vectors of the flexure mode are eventually the sinuous-shaped ripples in the out-of-plane direction. The energy of the flexure mode is $E_{n}\propto (n/L)^{2}$, where $L$ is the system length and $n$ is the mode index. This phonon energy can be regarded as the formation energy for the ripple in the monolayer MoS$_{2}$. The first flexure mode ($n=1$) has the lowest phonon mode energy, so it is the easiest one to be excited when the monolayer MoS$_{2}$ is compressed. The contour for $\omega=0$ is shown in Fig.~\ref{fig_dispersion_strain}~(a) as the green circle. The contour looks like a pair of glasses. All flexure modes with wave vectors inner the $\omega=0$ contour have negative frequency. The area enclosed by the $\omega=0$ contour reflects the overall instability of the monolayer MoS$_{2}$. The maximum wave vector on the contour is denoted by $k_{c}$. It gives a critical length of $L_{c}=2\pi/k_{c}$. It means that, for the particular uniaxial compression $\epsilon=-0.08$, the monolayer MoS$_{2}$ becomes unstable and forms a ripple, if its length is longer than $L_{c}$. In other words, a monolayer MoS$_{2}$ of length $L_{c}$ becomes unstable and is rippled, if it is uniaxially compressed by a strain stronger than 0.08.

Fig.~\ref{fig_dispersion_strain}~(b) shows the $\omega=0$ contour for monolayer MoS$_{2}$, which is compressed by various uniaxial strain. For curves from inside to outside, the uniaxial compression is $\epsilon=-0.01$, -0.02, -0.05, -0.08, -0.11, and -0.14. The contour for $\omega=0$ expands with increasing compression strength. It means that the critical wave vector, $k_{c}$, increases with increasing uniaixal compression, so the critical length decreases. In other words, a stronger compression is in need to generate ripples in a shorter monolayer MoS$_{2}$. This is quite reasonable considering larger formation energy for a ripple in shorter monolayer MoS$_{2}$.

Fig.~\ref{fig_dispersion_strain}~(c) shows the band structure for the flexure mode in monolayer MoS$_{2}$ under biaxial compression. The compression strain is $\epsilon=-0.08$. The negative frequency region also occurs and it is a circular area. Fig.~(d) shows the expansion of the expansion of the $\omega=0$ contour with increasing biaxial compression strength, $\epsilon=-0.01$, -0.02, -0.05, -0.08, -0.11, and -0.14. The six-fold symmetry in the contour is well kept due to the symmetric compression style.

From Fig.~\ref{fig_dispersion_strain}, we obtain the relationship between the critical length, $L_{c}$, and the critical compression value, $\epsilon_{c}$. This relationship is displayed in Fig.~\ref{fig_lc} for uniaixal compression (filled triangles) and for biaixal compression (filled squares). This relation provides us two pieces of equivalent information. On the one hand, a monolayer MoS$_{2}$ of length $L_{c}$ becomes unstable if it is compressed by a strain stronger than $\epsilon_{c}$. On the other hand, under a particular compression $\epsilon_{c}$, the monolayer MoS$_{2}$ becomes unstable if it is longer than $L_{c}$. The linear fitting shows that $L_{c}\propto \epsilon^{-0.5}$ in the small compression region. This dependence is exactly the same as the critical compression predicted by the classical Euler buckling of thin plates.\cite{TimoshenkoS1987} According to the Euler buckling theory, the critical compression for a thin plate under the uniaxial compression is related to its bending modulus $D$ and the in-plane tension stiffness $C_{11}$ through following formula,
\begin{eqnarray}
L_{c} = \sqrt{\frac{4\pi^2 D}{C_{11}}}  \sqrt{\frac{1}{|\epsilon_{c}|}}.
\label{eq_euler}
\end{eqnarray}
The Stillinger-Weber potential gives a bending modulus\cite{JiangJW2013bend}  $D=9.61$~{eV} and the in-plane tension stiffness\cite{JiangJW2013sw} $C_{11}=139.5$~{Nm$^{-1}$} for the monolayer MoS$_{2}$. Using these two quantities, we get the relation, $L_{c}=6.6|\epsilon_{c}|^{-0.5}$. For the uniaixal compression, Fig.~\ref{fig_dispersion_strain} shows that $L_{c}=6.5|\epsilon_{c}|^{-0.5}$, where the fitting coefficient of 6.5 is in excellent agreement with the result of 6.6 predicted by the Euler buckling formula.
\begin{figure}[htpb]
  \begin{center}
    \scalebox{1}[1]{\includegraphics[width=8cm]{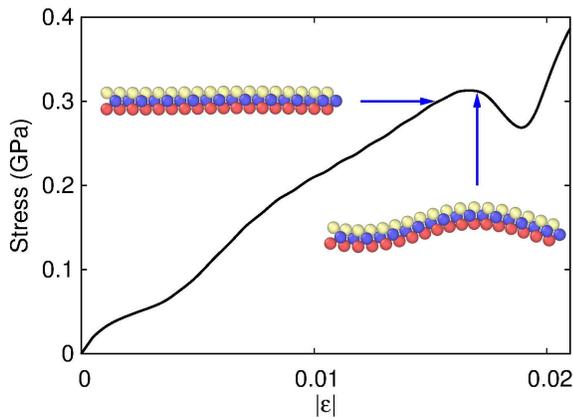}}
  \end{center}
  \caption{(Color online) The stress-strain relation during the biaxial compression of the monolayer MoS$_{2}$ with length $L=50$~{\AA} at 1.0~K. The instability happens at $\epsilon_{c}=-0.0169$. Insets show the occurrence of a ripple in the monolayer MoS$_{2}$ for compression above $\epsilon_{c}$.}
  \label{fig_stress_strain}
\end{figure}
 For large compression, our calculation results deviate from the linear Euler buckling theory, as the nonlinear effect becomes important in the large strain region. Fig.~\ref{fig_lc} shows that the biaxial compression is more effective than the uniaxial compression in the generation of ripples in the monolayer MoS$_{2}$. That is, for a monolayer MoS$_{2}$ of length $L_{c}$, the critical strain for the biaixal compression is smaller than that for the uniaxial compression.

We have now been aware that the monolayer MoS$_{c}$ becomes unstable under compression, and we have obtained, from the phonon band structure, the relationship between the critical length and the critical compression value. This relationship between the critical length and the critical compression can also be extracted from the MD simulation of the behavior for the monolayer MoS$_{2}$ under compression. We thus perform MD simulation for the structure evolution of the monolayer MoS$_{2}$ of length $L_{c}$ under compression. The widths of all simulated monolayer MoS$_{2}$ are 20~{\AA}. Periodic boundary conditions are applied in the two in-plane directions, and the free boundary condition is applied in the out-of-plane direction.

Fig.~\ref{fig_stress_strain} shows the stress-strain relation for the biaxial compression of the monolayer MoS$_{2}$ of 50~{\AA} in length. Two insets show the configuration just before and after the occurrence of the compression-induced ripple. The instability happens at $\epsilon_{c}=-0.0169$. This instability is actually the buckling of the monolayer MoS$_{2}$. A ripple is formed in the monolayer MoS$_{2}$ to accommodate the compression strain energy inside the system. The shape of this ripple is exactly the vibration morphology of the first flexure mode in the monolayer MoS$_{2}$ with periodic boundary conditions in both in-plane directions, which is also the predicted buckling mode in the Euler buckling theory.\cite{TimoshenkoS1987} From this simulation, we get the critical compression $\epsilon_{c}=-0.0169$ corresponding to a critical length $L_{c}=50$~{\AA}. We perform similar MD simulations for monolayer MoS$_{2}$ of different length. Both uniaixal and biaxial compressions are considered in the MD simulation. The relationship, $L_{c}$ v.s $\epsilon_{c}$, from the MD simulation is shown in Fig.~\ref{fig_lc} for uniaixal compression (open triangles) and for biaxial compression (open squares). The MD simulation results overall agree with results from the phonon band structure calculation. In particular, the MD simulation results confirm that the biaxial compression is more effective than the uniaxial compression for the generation of ripples in the monolayer MoS$_{2}$.

\jwj{We note that a similar energy gap also exists in the phonon dispersion of other dichalcogenides,\cite{DamnjanovicM2008mmp,SanchezAM} which have quite similar atomic configurations. Furthermore, these materials are close to each in their chemical properties, i.e they have similar covalent chemical bonding. Based on the above discussion, the strain effect on the phonon band gap is expected to be general in these dichalcogenides.}

\section{Conclusion}
In conclusion, we have performed both phonon band structure calculation and MD simulations to investigate the tension effect on the phonon bandgap and the compression induced instability (buckling) of the monolayer MoS$_{2}$. More specifically, we find that the uniaxial tension can effectively narrow the phonon bandgap between the acoustic and optical branches, and the bandgap is completely closed by applying a uniaxial tension of 0.145. The biaxial tension only has limited effect on the bandgap of the monolayer MoS$_{2}$. Hence, the uniaxial tension is expected to be an efficient approach for the manipulation of phonon related properties in the monolayer MoS$_{2}$. Based on the phonon band structure analysis and MD simulations, we demonstrate the instability and ripple formation in the monolayer MoS$_{2}$ under compression, where the relationship between the critical length and critical compression is in excellent agreement with the Euler buckling theory. The biaxial compression is found to be more effective than the uniaxial compression in the formation of ripples within the monolayer MoS$_{2}$.

\textbf{Acknowledgements} The work is supported by the Recruitment Program of Global Youth Experts of China and the start-up funding from the Shanghai University.

%
\end{document}